\documentclass[prl,superscriptaddress,twocolumn,epsf]{revtex4}
\usepackage{epsfig}

\newcommand{\braket}[2]{\langle #1 \,|\, #2 \rangle}
\newcommand{\ket}[1]{| \, #1 \rangle}
\newcommand{\bra}[1]{ \langle #1 \,  |}

\begin{document}
\title{Playing games in quantum mechanical settings: A necessary and sufficient condition}
\author{Junichi Shimamura}
\affiliation{SORST Research Team for Interacting Carrier
Electronics, CREST Research Team for Photonic Quantum Information,
Graduate School of Engineering Science, Osaka University, 1-3
Machikaneyama, Toyonaka, Osaka 560-8531, Japan}
\author{\c{S}ahin Kaya \"Ozdemir}
\email{ozdemir@qi.mp.es.osaka-u.ac.jp} \affiliation{SORST Research
Team for Interacting Carrier Electronics, CREST Research Team for
Photonic Quantum Information, Graduate School of Engineering
Science, Osaka University, 1-3 Machikaneyama, Toyonaka, Osaka
560-8531, Japan}
\author{Nobuyuki Imoto}
\affiliation{SORST Research Team for Interacting Carrier
Electronics, CREST Research Team for Photonic Quantum Information,
Graduate School of Engineering Science, Osaka University, 1-3
Machikaneyama, Toyonaka, Osaka 560-8531, Japan} \pagestyle{plain}
\pagenumbering{arabic} \maketitle {\small {\bf A number of recent
studies
\cite{Eisert1,Meyer,Weber,Ben,Du2,Flitney,Shima,Ozdemir,Lee} have
focused on novel features in game theory when the games are played
using quantum mechanical toolbox (entanglement, unitary operators,
measurement). Researchers have concentrated in two-player-two
strategy, $2\times 2$, dilemma containing classical games, and
transferred them into quantum realm  showing that in quantum pure
strategies dilemmas in such games can be resolved if entanglement
is distributed between the players armed with quantum operations.
Moreover, it became clear that the players receive the highest sum
of payoffs available in the game, which are otherwise impossible
in classical pure strategies. Encouraged by the observation of
rich dynamics of physical systems with many interacting parties
and the power of entanglement in quantum versions of $2\times 2$
games, it became generally accepted that quantum versions can be
easily extended to $N$-player situations by simply allowing
$N$-partite entangled states. In this article, however, we show
that this is not generally true because the reproducibility of
classical tasks in quantum domain imposes limitations on the type
of entanglement and quantum operators. We propose a benchmark for
the evaluation of quantum and classical versions of games, and
derive the necessary and sufficient conditions for a physical
realization. We give examples of entangled states that can and
cannot be used, and the characteristics of quantum operators used
as strategies.}

Mathematical models and techniques of game theory have
increasingly been used by computer and information scientists,
i.e., distributed computing, cryptography, watermarking and
information hiding tasks can be modelled as games
\cite{Moulin,Cohen,Conway,Shen,Ettinger,Pateux,Luiz}. Therefore,
new directions have been opened in the interpretation and use of
game theoretical toolbox which has been traditionally limited in
economical and evolutionary biology problems \cite{Books}. This is
not a surprise because there is a very strong connection between
the two: Information \cite{Lee1}. Game theory deals with
situations where players make decisions, and then depending on
their decisions, the outcome of games are determined. This process
can be modelled as the flow of information. Since generation,
transmission, storage, manipulation and processing of information
need physical means, information is governed by the laws of
physics. Therefore, information is closely related to physics and
hence to quantum mechanics. In short, information is the common
link among game theory, physics, quantum mechanics, computation
and information sciences. Along this line of thinking, researchers
introduced the quantum mechanical toolbox into game theory to see
what new features will arise combining these two beautiful areas
of science.

Quantum mechanics is introduced into game theory through the use
of quantum bits (qubits) instead of classical bits, entanglement
which is a quantum correlation with a highly complex structure and
is considered to be the essential ingredient to exploit the
potential power of quantum information processing, and the quantum
operations.  This effort, although has been criticized on the
basis of using artificial models \cite {Ben1,Enk}, has produced
significant results: (i) Dilemmas in some games can be resolved
\cite{Eisert,Eisert1,Ozdemir,Shima,Simon,Adrian}, (ii) playing
quantum games can be more efficient in terms of communication
cost; less information needs to be exchanged in order to play the
quantized versions of classical games \cite{Lee,Lee1,Ozdemir}, and
(iii) entanglement is not necessary for the emergence of Nash
Equilibrium but for obtaining the highest possible sum of payoffs
\cite{Ozdemir}, and (iv) Quantum advantage does not survive in the
presence of noise above critical level \cite{Johnson3,Ozdemir}. In
addition to these fundamental results and efforts, Piotrowski and
Sladkowski have described, in a series of papers, market
phenomena, bargaining, auction and finance using quantum game
theory \cite{Edward1,Edward2,Edward3}. The positive results are
consequences of the fact that sharing entanglement, using quantum
operators and measurements allow players to have a greater number
of strategies to choose from when compared to the situation in
classical games.

In this paper, we focus on (i) the extent of entangled states and
quantum operators that can and cannot be used in multi-player
games, and (ii) comparison of the results of classical games and
their quantized versions on a fair basis by introducing a
benchmark. Moreover, this study attempts to clarify a relatively
unexplored area of interest in quantum game theory, that is the
effects of different types of entangled states and their use in
multi-player multi-strategy games in quantum settings. Our
approach, which will become clear in the following, to these
points are based on the reproducibility of classical games in the
physical schemes used for the implementation of their quantized
versions.

We should mention that reproducibility requires that a chosen
model of game should simulate both quantum and classical versions
of the game to make a comparative analysis of quantum and
classical strategies, and to discuss what can or cannot be
attained by introducing quantum mechanical toolbox. This is indeed
what has been observed in quantum Turing machine (QTM). A QTM can
simulate the classical Turing machine (CTM) under special
conditions, and can reproduce the results of the original CTM.
Therefore, we think the reproducibility criterion must be taken
into consideration whenever a comparison between classical and
quantum versions of a task is needed. An important consequence of
this criterion in game theory is the main contribution of this
study: The derivation of the necessary and sufficient condition
for entangled states and quantum operators that can be used in the
quantized versions of classical games.

{\bf Definitions and models:} We start by introducing some basic
definitions and the model of the quantized classical game that is
considered in this study. In classical game theory, a strategic
game is defined by $\Gamma=[N,(S_{i})_{i\in N},(\$_{i})_{i\in N}]$
where $N$ is the set of players, $S_{i}=\{s_{i}^{1},s_{i}^{2},
\ldots , s_{i}^{m}\}$ is the set of pure strategies available to
the $i$-th player with $m$ being the number of strategies, and
$\$_{i}$ is the payoff function for the $i$-th player. When the
strategic game $\Gamma$ is played with pure strategies, the $i$-th
player chooses one of the strategies from the set $S_i$. With all
players applying a pure strategy (each player chooses only one
strategy from the strategy set), the joint strategy of the players
is denoted by $\vec{s}_k=(s_{1}^{l_{1}},s_{2}^{l_{2}}, \ldots,
s_{N}^{l_{N}})$ with $l_{i}\in \{ 1,2,3, \ldots ,m \}$ and
$k=\sum_{i=1}^{N}(l_i -1)m^{i-1}$. Then the $i$-th player's payoff
function is represented by $\$_i (\vec{s}_k)$ when the joint
strategy set $\vec{s}_k$ is chosen, i.e., the payoff functions of
all players corresponding to the unique joint strategy $\vec{s}_k$
can be represented by  $\vec{\$}=(\$_1 (\vec{s}_k),\$_2
(\vec{s}_k), \ldots ,\$_{N} (\vec{s}_k))$ and it is uniquely
determined from the payoff matrix of the game. Players may choose
to play with mixed strategies, that is they randomize among their
pure strategies resulting in the expected payoff
\begin{equation} \label{mix:1} F_i(q_1,\cdots,q_N) = \sum_{s^{l_1}_1
\in S_1}\cdots \sum_{s^{l_N}_N \in S_N}\left\{\prod_{j=1}^{N} q_j
(s^{l_j}_j)\right\} f_i(s^{l_1}_1,\cdots,s^{l_N}_N)
\end{equation}
where $q_j(s^{l_j}_j)$ represents the probability that $j$-th
player chooses the pure strategy $s^{l_j}_j$ and $f_i$ is the
corresponding payoff function for the $i$-th player.

Most of the studies on quantum versions of classical games have
been based on the model proposed by Eisert {\it et al.}
\cite{Eisert1}. In this model, the strategy set of the players
consists of unitary operators which are applied locally on a
shared entangled state by the players. A measurement by a referee
on the final state after the application of the operators maps the
chosen strategies of the players to their payoff functions. In
this model, the two strategies of the players in the original
classical game is represented by two unitary operators,
$\{\hat{\sigma}_0$, $i\hat{\sigma}_{y}\}$, i.e., in Prisoner's
Dilemma $\{\hat{\sigma}_0$ and $i\hat{\sigma}_{y}\}$ respectively
corresponds to ``Cooperate" and ``Defect".

In this study, however, we consider the following model of a
quantum version of classical games for $N$-player-two-strategy
games, which is more general than Eisert {\it et al.}'s model
\cite{Eisert1} and includes it. In our model \cite{Shima1}, (i) A
referee prepares an $N$-qubit entangled state $\ket{\Psi}$ and
distributes it among $N$ players, one qubit for each player. (In
order to see features intrinsic to quantumness, we focus on a
shared entangled state among the players, and exclude the trivial
case where a product state is distributed.) (ii) Each player
independently and locally applies a unitary operator chosen from
the SU(2) set on his qubit, i.e, the $i$-th player applies
$\hat{u}_i$. (We restrict ourselves to the entire set of SU(2)
because the global phase is irrelevant). Hence, the combined
strategies of all the players is represented by the tensor product
of all players' unitary operators as $\hat{x}=\hat{u}_1 \otimes
\hat{u}_2 \otimes \cdots \otimes \hat{u}_N$, which generates the
output state $\hat{x}\ket{\Psi}$ to be submitted to the referee.
(iii) Upon receiving this final state, the referee makes a
projective measurement $\{ \Pi_j \}_{j=1}^{2^{N}}$ which outputs
$j$ with probability ${\rm Tr}[\Pi_{j}
\hat{x}\ket{\Psi}\bra{\Psi}\hat{x}^{\dagger}]$, and assigns
payoffs chosen from the payoff matrix, depending on the
measurement outcome $j$. Therefore, the expected payoff of the
$i$-th player is described by
\begin{eqnarray}
&&\hspace{-5mm}F_{i}(\hat{U}_1,\cdots,\hat{U}_N)=\nonumber \\
&&\hspace{-5mm}{\rm Tr}\left[\left(\sum_j
a_{j}^i\Pi_{j}\right)\left(\hat{U}_1\otimes \cdots \otimes
\hat{U}_N \ket{\Psi_{\rm in}}\bra{\Psi_{\rm
in}}{\hat{U}_1}^{\dagger}\otimes \cdots \otimes
{\hat{U}_N}^{\dagger}\right)\right]\nonumber\\\label{eq:quantum}
\end{eqnarray}
where $\Pi_j$ is the projector and $a_j^{i}$ is the $i$-th
player's payoff when the measurement outcome is $j$. This model
can be implemented in a physical scheme with the current level of
experimental techniques and technology of quantum mechanics.

{\bf Reproducibility criterion:} In the following, reproducibility
criterion corresponds to the reproducibility of a multi-player
two-strategy classical game in the quantization model explained
above. This criterion requires that the expected payoff given in
eq. (\ref{mix:1}) is reproduced in the quantum version, too
\cite{Shima1}.

First, we consider the reproducibility problem only in pure
strategies, $q_j(s^{l_j}_j)=1$ for all $j=1\cdots N$, (conditions
for the reproducibility including the mixed strategies will be
discussed later below). In this model, we require that a classical
game be reproduced when each player's strategy set is restricted
to two unitary operators, $\{ \hat{u}_{i}^{1},\hat{u}_{i}^{2} \}$,
corresponding to the two pure strategies in the classical game.
When the classical game is played in this model, the combined pure
strategy of the players is represented by
$\hat{x}_k=\hat{u}_{1}^{l_1} \otimes \hat{u}_{2}^{l_2} \otimes
\cdots \otimes \hat{u}_{N}^{l_N}$ with $l_i=\{1,2\}$ and
$k=\sum_{i=1}^{N}(l_i -1)2^{i-1}$. Thus the output state becomes
$\ket{\Phi_k}=\hat{x}_k\ket{\Psi}$ with $k = \{1,2, \ldots, 2^N
\}$. When the strategy combination $\hat{x}_{k}$ is played, Eq.
(\ref{eq:quantum}) becomes
\begin{eqnarray}
F_{i}(\hat{u}_1^{l_1},\ldots,\hat{u}_N^{l_N}) &=& {\rm
Tr}\left[\left(\sum_j a_{j}^i\Pi_{j}\right)\hat{x}_k \ket{\Psi}
\bra{\Psi}\hat{x}_{k}^{\dagger}\right]
\nonumber \\
&=& \sum_j a_{j}^i{\rm Tr}[\Pi_{j}\hat{x}_k \ket{\Psi}
\bra{\Psi}\hat{x}_k^{\dagger} ]
\end{eqnarray}
where the measurement outcome $j$ occurs with probability ${\rm
Tr}[\Pi_{j}(\hat{x}_k \ket{\Psi} \bra{\Psi}\hat{x}_{k}^{\dagger}
]$. Playing with pure strategies requires referee discriminate all
the possible output states $\ket{\Phi_k}$ deterministically in
order to assign payoffs uniquely. That is, the projector $\{ \Pi_j
\}_{j=1}^{2^{N}}$ has to satisfy ${\rm Tr}[\Pi_{j}
\ket{\Phi_k}\bra{\Phi_k}]=\delta_{jk}$, which is possible if and
only if
\begin{equation}
\braket{\Phi_\alpha}{\Phi_\beta}=\delta_{\alpha\beta} \;\; \forall
\alpha, \beta. \label{condition}
\end{equation}
Under this distinguishability condition,  we see that
$F_{i}(\hat{u}_1^{l_1},\ldots,\hat{u}_N^{l_N})=a_k^i =
f_i(s_{l_1},...,s_{l_N})$. Therefore, Eq. (\ref{condition})
becomes the {\it necessary condition} for the reproducibility of
classical games in the quantum model. Imposing this necessary
condition on several multiparty-entangled states, we have found
\cite{Shima1}: (a) Bell states and any two-qubit pure state
satisfy it if the two unitary operators for two players are chosen
as $\{\hat{\sigma}_0,\hat{\sigma}_{x} \}$ and $\{ \hat{\sigma}_0,
i\hat{\sigma}_{y} \}$, respectively. (b) Multipartite GHZ-like
states of the form $(\ket{00 \ldots 0} + i \ket{11 \ldots
1})/\sqrt{2}$ satisfy the above condition if the unitary operators
of the players are chosen as $\{ \hat{\sigma}_0, i\hat{\sigma}_y
\}$. Entangled states that can be obtained from GHZ state by local
unitary transformations also satisfy it. (c) $N$-party form of the
W state, defined as $|W_{N}\rangle=|N-1,1\rangle/ {\sqrt{N}}\, (N
\ge 3)$, where $|N-1,1\rangle$ is a symmetric state with $N-1$
zeros and $1$ one, e.g.
$|2,1\rangle=|001\rangle+|010\rangle+|100\rangle$, does not
satisfy it, therefore the entangled state $|W_{N}\rangle$ cannot
be used in this model of quantum games. (d) Among the Dicke
states, which is a class of symmetric states represented as
$\ket{N-m,m}/\sqrt{{}_{N}C_m}$ with $(N-m)$ zeros and $m$ ones
(${}_{N}C_m$ denoting the binomial coefficient), only the states
$\ket{1,1}$ and $\ket{2,2}$ satisfy the distinguishability
condition.

{\bf Quantum operators and distinguishability condition:} In the
following, we will discuss some basic properties of quantum
operators which satisfy the distinguishability condition, and show
that this condition is also the sufficient condition for the
reproducibility of classical games.

Let us assume that for a given entangled state, $\ket{\Psi}$,
satisfying the distinguishability condition, we find two unitary
operators corresponding to the classical pure strategies as
required in the model proposed above for each player. Moreover,
considering that only the first player changes his operator while
the others stick to their first operator, we obtain $\ket{\Phi_0}=
\hat{u}_1^1 \otimes \hat{u}_2^1 \otimes \cdots \otimes
\hat{u}_N^1\ket{\Psi}$ and $\ket{\Phi_1}= \hat{u}_1^2 \otimes
\hat{u}_2^1 \otimes \cdots \otimes \hat{u}_N^1\ket{\Psi}$.
Imposing the distinguishability criterion on this simple case, we
arrive at the condition,
\begin{equation}
\bra{\Psi} (\hat{u}_1^1)^{\dagger}\hat{u}_1^2 \otimes \hat{I}
\otimes \cdots \otimes \hat{I} \ket{\Psi} = 0. \label{eq1:con}
\end{equation}
Since $(\hat{u}_1^1)^{\dagger}\hat{u}_1^2$ is a normal operator,
it can be diagonalized by a unitary operator $\hat{z}_1$.
Furthermore, since $(\hat{u}_1^1)^{\dagger}\hat{u}_1^2$ is a SU(2)
operator, the eigenvalues are given by $e^{i \phi_ 1}$ and $e^{-i
\phi_1}$. Then, we can transform eq. (\ref{eq1:con}) into
\begin{eqnarray}
&&\hspace{-15mm}\bra{\Psi}\, \hat{z}_1^{\dagger} \,
(\hat{z}_1(\hat{u}_1^1)^{\dagger}\hat{u}_1^2 \hat{z}_1^{\dagger})
\, \hat{z}_1 \otimes
\hat{I} \otimes \cdots \otimes \hat{I} \,\ket{\Psi} \nonumber \\
&&\hspace{-10mm} =  \bra{\Psi'} \left[
\begin{array}{cc}
e^{i \phi_1 } & 0 \\
0   & e^{-i \phi_1}\\
\end{array}
\right]\otimes \hat{I} \otimes \cdots \otimes \hat{I} \ket{\Psi '}
=0, \label{eq2:con}
\end{eqnarray}
where $\ket{\Psi'}= \hat{z}_1 \otimes \hat{I} \otimes \cdots
\otimes \hat{I} \ket{\Psi}$. We write the state $\ket{\Psi'}$ on
computational basis as $\ket{\Psi'} = \sum_{i_j \in \{0,1\}}
c_{i_1 \, i_2 ... i_N} \ket{i_1}\ket{i_2}\cdots\ket{i_N}$ and
substitute this into Eq. (\ref{eq2:con}). After some
straightforward matrix and trigonometric manipulations we obtain
\begin{eqnarray}
&&\hspace{-10mm}\bra{\Psi'} \left[
\begin{array}{cc}
e^{i \phi_1 } & 0 \\
0   & e^{-i \phi_1}\\
\end{array}
\right]\otimes \hat{I} \otimes \cdots \otimes \hat{I}
\ket{\Psi '} \nonumber \\
&=& e^{i \phi_1}  \sum_{i_j \in \{0,1\}} |c_{0 \, i_2 ... i_N}|^2
+ e^{- i \phi_1} \sum_{i_j \in \{0,1\}} |c_{1 \, i_2 ... i_N}|^2
\nonumber
\\
&=& \cos \phi_1 + i \left(2\sum_{i_j \in \{0,1\}} |c_{0 \, i_2 ...
i_N}|^2 - 1\right)\sin \phi_1 = 0.
\end{eqnarray}
In order for the above equality to hold, $\cos\phi_1 = 0$ and
$2\sum_{i_j \in \{0,1\}} |c_{0 \, i_2 ... i_N}|^2 - 1=0$ must be
satisfied. The equation $\cos\phi_1 = 0$ implies that the
diagonalized form $\hat{D}_1 = \hat{z}_1(
\hat{u}_1^1)^{\dagger}\hat{u}_1^2 \hat{z}_1^{\dagger}$ can be
written as
\begin{equation}
\hat{D}_1 = \left[
\begin{array}{cc}
i & 0 \\
0 & -i \\
\end{array}
\right].\label{eq8:dia}
\end{equation}
This argument holds for all players, therefore we write $\hat{z}_k
(\hat{u}_k^1)^{\dagger}\hat{u}_k^2 \hat{z}_k^{\dagger} =\hat{D}_k
=i\hat{\sigma}_z$ for $k=1,\cdots,N$. For example, let us consider
the case of the four-party Dickie State $\ket{2,2}$, which
satisfies the distinguishability criterion with the unitary
operators $\hat{u}_{k=1,2,3,4}^1=\hat{I}$,
$\hat{u}_{k=1,2,3}^2=i(\sqrt{2}\hat{\sigma}_{z}+\hat{\sigma}_{x})/\sqrt{3}$,
and $\hat{u}_{4}^2=i\hat{\sigma}_{y}$. It can easily be verified
that the eigenvalues of  $\hat{u_k}^{1\dagger}\hat{u}_k^2$ for all
players are $i$ and $-i$, and they are already in the diagonalized
form of Eq. (\ref{eq8:dia}). For the GHZ state, all players should
have the operators $\hat{u}_{k}^1=\hat{I}$ and
$\hat{u}_{k}^1=i\hat{\sigma}_y$ is the same as the operator set of
the fourth player of $\ket{2,2}$, and therefore can be written as
in Eq. (\ref{eq8:dia}).

Next we consider the following scenario: Each player has two
operators satisfying the above properties. Instead of choosing
either of these operators, they prefer to use a linear combination
of their operator set. Let this operator be $\hat{w}_k = \cos
\theta_k \hat{u}_k^1 + \sin \theta_k \hat{u}_k^2$ for the $k$-th
player. Then, we ask the questions (i) Does the property of the
operators $\hat{u}_k^1$ and $\hat{u}_k^2$ derived from the
distinguishability condition impose any condition on the operator
$\hat{w}_k$?, and (ii) What does the outcome of the game played in
the quantum version with the operator $\hat{w}_k$ imply? Since
$\hat{z}_k (\hat{u}_k1)^{\dagger}\hat{u}_k2\hat{z}_k^{\dagger}$ is
in the diagonalized form $\hat{D}$, we can write $\hat{w}_k$ in
such a way that it contains $\hat{D}$. In order to do this, we
look at the operator $\hat{w}_k^{\dagger}\hat{w}_k$ which is given
as
\begin{eqnarray}\label{eq1970:unitw}
\hat{w}_k^{\dagger} \hat{w}_k &=&
(\hat{u}_k^{1\dagger}\cos\theta_{k}
+ \hat{u}_k^{2\dagger}\sin\theta_{k})
(\hat{u}_k^1 \cos\theta_{k} +
\hat{u}_k^2\sin\theta_{k}) \nonumber \\
&=& \hat{I} + \cos\theta_k \sin\theta_k
(\hat{u}_k^{1\dagger}\hat{u}_k^2
+ \hat{u}_k^{2\dagger}\hat{u}_k^1) \nonumber \\
&=& \hat{I} + \cos\theta_k \sin\theta_k (\hat{z}_k^{\dagger}
\hat{D}\hat{z}_k
+ \hat{z}_k^{\dagger} \hat{D}^{\dagger} \hat{z}_k  )
\nonumber \\
&=& \hat{I} + \cos\theta_k \sin\theta_k (\hat{z}_k^{\dagger}
\hat{D}\hat{z}_k
-  \hat{z}_k^{\dagger} \hat{D} \hat{z}_k  ) \nonumber \\
&=& \hat{I},
\end{eqnarray} where we have used
$\hat{u}_k^{1\dagger}\hat{u}_k^2 = \hat{z}_k^{\dagger}
\hat{D}\hat{z}_k$, and $\hat{D}^{\dagger}= - \hat{D}$ since
$\hat{D}$ is anti-hermitian. Therefore, as seen in Eq.
(\ref{eq1970:unitw}), the distinguishability condition requires
that $\hat{w}_k$ be a unitary operator. In order to find the
outcome of the game when players use the operators $\hat{w}_k=\cos
\theta_k \hat{u}_k^1 + \sin \theta_k \hat{u}_k^2$, we substitute
$\hat{w}_k$ into Eq. (\ref{eq:quantum}) and obtain
\begin{widetext}
\begin{eqnarray}
F_{k}(\hat{w}_1,\ldots,\hat{w}_N) &=& a_1^{k}\,{\cos}^2{\theta_1}
{\cos}^2{\theta_2}\cdot \cdot{\cos}^2{\theta_N} +
a_2^{k}\,{\sin}^2{\theta_1} {\cos}^2{\theta_2}\cdot
\cdot{\cos}^2{\theta_N} + \cdots + a_{2^N}^{k}\,{\cos}^2{\theta_1}
{\sin}^2{\theta_2}\cdot \cdot{\sin}^2{\theta_N} \nonumber \\
&=& a_1^{k}\,p_1 p_2 \cdot \cdot p_N + a_2^{k} \,(1-p_1)p_2 \cdot
\cdot p_N + \cdots + a_{2^N}^{k} \, (1-p_1)(1-p_2)\cdot \cdot(1-p_N)
\nonumber \\
&=& \sum_{s_1 \in S_1} .. \sum_{s_N \in S_N} \left( \prod \,
q_1(s_1) \ldots q_N(s_N) \right) \, f_k(s_1,..,s_N),
\label{eq:reproduce}
\end{eqnarray}
\end{widetext} where $p_k = \cos2 \theta_k$ represents the
probability that $k$-th player chooses the strategy $s_{1_k} \in
S_k$. We can see that Eq. (\ref{eq:reproduce}) has the same form
of the expected payoff given in Eq. (\ref{mix:1}) for the
classical game implying that when players choose $\hat{w}_k = \cos
\theta_k \hat{u}_k + i \sin \theta_k \hat{v}_k$ as their
strategies, the payoff for the mixed strategies in the classical
games is reproduced in this quantum version. Therefore, we
conclude that the distinguishability condition of Eq.
(\ref{condition}) is the {\it necessary and sufficient condition}
for the reproducibility of a classical game in the quantum
version. This is because when players apply their pure strategies
with unit probabilities, results of classical pure strategy, and
when they apply a linear combination of their pure strategies
results of classical mixed strategy are reproduced in the quantum
setting.

{\bf Entangled states and distinguishability condition:} After
stating the properties of operators which satisfy the
distinguishability criterion, we proceed to investigate the
properties of the class of entangled states which satisfy the
distinguishability condition.

Suppose that an N-qubit state $\ket{\Psi}$ and two unitary
operators $\{\hat{u}_k, \hat{v}_k\}$ satisfy the
distinguishability criterion. Let us consider the
distinguishability criterion between the states $\ket{\Phi_0}=
\hat{u}_1 \otimes \hat{u}_2 \otimes \cdots \otimes \hat{u}_N
\ket{\Psi}$ and $\ket{\Phi_1}= \hat{v}_1 \otimes \hat{u}_2 \otimes
\cdots \otimes \hat{u}_N\ket{\Psi}$. Using the properties of the
operators derived above, the distinguishability criterion for
these two states is written as
\begin{eqnarray}
&&\hspace{-25mm}\bra{\Psi} \hat{z}_1^{\dagger} \,
\hat{z}_1\hat{u}_1^{\dagger} \hat{v}_1 \hat{z}_1^{\dagger} \,
\hat{z}_1\otimes \hat{I} \otimes \cdots \otimes
\hat{I} \ket{\Psi} \nonumber \\
&=&  \bra{\Psi'}\hat{D}_1 \otimes \hat{I} \otimes \cdots \otimes
\hat{I} \ket{\Psi'} =0, \label{eq3:con}
\end{eqnarray}
where $\hat{z}_1$ is a unitary operator diagonalizing
$\hat{u}_1^{\dagger} \hat{v}_1$ and $\ket{\Psi'} = \hat{z}_1
\otimes \hat{I} \otimes \cdots \otimes \hat{I} \ket{\Psi}$. This
implies that if the N-qubit state $\ket{\Psi}$ and the operators
$\{\hat{u}_k, \hat{v}_k\}$ satisfy the distinguishability
criterion, then the state $\ket{\Psi '} = \hat{z}_1 \otimes
\hat{z}_2 \otimes \cdots \otimes \hat{z}_N \ket{\Psi}$ and the
unitary operators $\{\hat{D}, \hat{I} \}$ should satisfy the
distinguishability criterion, too. Since the global phase is
irrelevant, Eq. (\ref{eq3:con}) can be further reduced to
\begin{equation}
\bra{\Psi'}\left[
\begin{array}{cc}
1 & 0  \\
0 & -1 \\
\end{array}
\right] \otimes \hat{I} \otimes \cdots \otimes \hat{I} \ket{\Psi'}
= 0.
\end{equation}
Thus, we end up with the following $2^{N}-1$ equalities to be
satisfied for the distinguishability criterion:
\begin{eqnarray}
&&\bra{\Psi'} \hat{\sigma}_z \otimes \hat{I}\otimes \hat{I}
\otimes \cdots
\otimes \hat{I} \ket{\Psi'} = 0, \nonumber \\
&& \bra{\Psi'} \hat{I} \otimes \hat{\sigma}_z \otimes \hat{I}
\otimes \cdots
\otimes \hat{I} \ket{\Psi'} = 0, \nonumber \\
&&  \;\;\;\;\;\;\;\;\;\; \vdots \nonumber \\
&&\bra{\Psi'} \hat{\sigma}_z \otimes \hat{\sigma}_z \otimes \cdots
\otimes \hat{\sigma}_z \ket{\Psi'} = 0.
\end{eqnarray}
When $\ket{\Psi'}$ is described as $\sum_{i_j \in \{ 0,1 \}}
c_{i_1 i_2 ... i_N} \ket{i_1}\ket{i_2}\cdots\ket{i_N}$, these
equations and the normalization condition can be written in the
matrix form as
\begin{equation}
\left[
\begin{array}{ccccc}
1 & 1 & \ldots & -1 & -1 \\
  &  & \ldots &   &     \\
  &   &  \vdots    &   &    \\
1 & 1 & \ldots & 1 & 1
\end{array}
\right] \left[
\begin{array}{l}
|c_{00...0}|^2 \\
|c_{00...1}|^2 \\
  \;\;\;\; \vdots \\
|c_{11...1}|^2 \\
\end{array}
\right] = \left[
\begin{array}{l}
0 \\
0 \\
\vdots \\
1 \\
\end{array}
\right], \label{eq:diagonal}
\end{equation}
where the last row is the normalization condition. The row vector
corresponds to the diagonal elements of $\hat{\sigma}_z^{\{0,1\}}
\otimes \cdots \otimes \hat{\sigma}_z^{\{0,1\}}$ where
$\hat{\sigma}_z^{0}$ is defined as $\hat{I}$. Here, let us
consider the operator $\hat{x},\hat{y} \in
(\hat{\sigma}_z^{\{0,1\}})^{\otimes N}$. The product of two
operators $\hat{x}\,\hat{y} $ also belongs to
$(\hat{\sigma}_z^{\{0,1\}})^{\otimes N}$. Since ${\rm
Tr}[\hat{\sigma}_z] = 0$, when $\hat{x} \ne \hat{y}$, ${\rm
Tr}[\hat{x}\,\hat{y}] = {\rm Tr}[\hat{x}]\,{\rm Tr}[\hat{y}] = 0$.
This means that every two row vectors are orthogonal with each
other, thus the matrix in Eq. (\ref{eq:diagonal}) has an inverse
and $|c_{i_1 i_2 ... i_N}|^2$ are determined uniquely. Since each
row but the last contains equal number of $1$ and $-1$, we can
easily find that $|c_{i_1 i_2 ... i_N}|^2= 1/N$. This implies that
if a state satisfies the distinguishability condition, then it
should be transformed by local unitary operators into the state
which contains all possible terms with the same magnitude but
different relative phases, i.e.,
\begin{equation}
\ket{\Psi'} = \sum_{i_j \in \{ 0,1 \}} \frac{1}{\sqrt{N}} e^{i
\phi_{i_1 i_2 .. i_N}}
\ket{i_1}\ket{i_2}\cdots\ket{i_N}.\label{eq15:st}
\end{equation} As examples for this case, let us consider
the product state and the GHZ state, which satisfy the
distinguishability criterion. We can easily find that the product
state is transformed into the form of the above state by Hadamard
operator, written as $(\hat{\sigma_x} + \hat{\sigma_z})/\sqrt{2}$.
The GHZ state is also transformed into the form of Eq.
(\ref{eq15:st}) by $(e^{i \frac{\pi}{4}} \hat{I} + e^{- i
\frac{\pi} {4}} \hat{ \sigma}_z + \hat{\sigma}_y )/\sqrt{2}$ for
one player and Hadamard operator for the others.

{\bf Reproducibility criterion as a benchmark:} This criterion
requires that in any physical model to play the quantum version,
it should be possible to play the classical game as well. It is
only when this is possible we can compare the outcomes of
classical and quantum versions to draw conclusions on whether the
quantum version has advantage over the classical one or not.
Therefore, the fist thing the physical scheme should provide is
the availability of unitary operators corresponding to classical
pure strategies. If there exists such operators then one can
compare the outcomes for the pure strategies. To make this point
clear, let us consider the entangled state $W$ for which one
cannot find two unitary operators $\{\hat{u}_k,\hat{v}_k\}$
satisfying the croterion. When any game is played using this
entangled state with unitary operators chosen from the SU(2) set
one cannot obtain the outcome of the classical game in pure
strategies. Moreover, the payoffs that will be obtained become a
probability distribution over the entries of the payoff matrix of
the classical game. Therefore, comparing the quantum version using
$W$ state with the classical game in pure strategies is not fair.
In the same way, comparing the quantum version played with GHZ and
$W$ states is not fair either because in GHZ the payoffs delivered
to the players are unique entries from the classical payoff table
for GHZ state, thus original classical game results in pure
strategies are reproduced which is not the case for $W$ state.
Therefore, we think that the reproducibility criterion constitutes
a benchmark not only for the evaluation of entangled states and
operators in quantum games but also in the comparative analysis of
classical games and their corresponding quantum versions.

{\bf Conclusion:} In this paper, for the first time, we give the
 necessary and sufficient  condition to play quantized version of classical
games in a physical scheme. This condition is introduced here as
the ''reproducibility criterion," or ''distinguishability
condition" and it provides a fair basis to compare quantum
versions of games with their classical counterparts. This
benchmark requires that results of the classical games be
reproduced in the model of the quantum version. The necessary and
sufficient condition we give here shows that a large class of
multipartite entangled states cannot be used in the quantum
version of classical games; and the operators that might be used
should have a special form in their diagonalized form. Given two
unitary operators $\{\hat{u}_k,\hat{v}_k\}$ corresponding to
classical pure strategies and satisfying the distinguishability
criterion, we can reproduce the results of classical games in pure
strategies in the physical scheme. Moreover, provided that the
players choose unitary operators in the space spanned by
$\hat{u}_k$ and $\hat{v}_k$, mixed strategy results of classical
games can be also reproduced in the quantum setting.

\begin{acknowledgments}
The authors thank M. Koashi, F. Morikoshi and T. Yamamoto for
helpful discussions and warm support during this research.
\end{acknowledgments}


}
\end{document}